\documentclass[amsmath,amssymb]{article}

\usepackage{mathtools,cancel}

\usepackage{soul,xcolor}
\usepackage{xcolor}
\usepackage{multirow}
\usepackage{pstricks}
\usepackage{dcolumn}
\usepackage{bm}
\usepackage{latexsym}
\usepackage{dcolumn}
\usepackage[utf8x]{inputenc} 
\usepackage{amsmath}
\usepackage{amsfonts,amssymb}
\usepackage{graphicx,epsfig}
\usepackage{color}
\usepackage{psfrag}
\usepackage{amsthm}
\usepackage{ulem} 
\usepackage{soul} 
\usepackage{flushend}
\usepackage{titlesec}
\usepackage{slashed}
\usepackage{authblk}
\usepackage[font=footnotesize]{caption}
\usepackage[hmarginratio=1:1,top=32mm,columnsep=60pt]{geometry}

\newcommand{\bea}{\begin{eqnarray}}
\newcommand{\eea}{\end{eqnarray}}
\newcommand{\be}{\begin{equation}}
\newcommand{\ee}{\end{equation}}
\newcommand{\ba}{\begin{array}}
\newcommand{\ea}{\end{array}}

\begin{document}

\unitlength = 1mm
\begin{flushright}
SHIP-HEP-2021-01
\end{flushright}

\title{The Dark $Z'$ and Sterile Neutrinos Behind Current Anomalies}  

\author{\normalsize A.~ Hammad$^{\dagger,\gamma}$, Ahmed Rashed$^{\delta}$ and S. Moretti$^{\ddagger\xi}$}
\affil{\small
$^\dagger$Institute of Convergence Fundamental Studies, Seoul National University of Science and Technology,  Seoul, 01811, Korea\\
$^\gamma$Centre for Theoretical Physics, the British University in Egypt, P.O. Box 43, Cairo 11837, Egypt\\
$^{\delta}$Department  of Physics,  Shippensburg University of Pennsylvania,\\
Franklin Science Center, 1871 Old Main Drive, Pennsylvania, 17257, USA\\
$^{}\ddagger$ School of Physics and Astronomy, University of Southampton, Highfield, Southampton SO17 1BJ, UK\\
$^{}\xi$ Department of Physics and Astronomy, Uppsala University,
Box 516, SE-751 20 Uppsala, Sweden

 }
\date{\today}

{\let\newpage\relax\maketitle}

\begin{abstract}
\noindent \normalsize 
We show how, in the $B-L$ extension of the SM (BLSM) with an Inverse Seesaw (IS) mechanism for neutrino mass generation, a light $Z'$ state with moderate couplings to SM objects, hence `dark' in its nature, can be associated, in conjunction with light sterile neutrinos,  to some present day data anomalies, such as the anomalous magnetic moment of the muon as well as    a possible signal indicating the existence of sterile neutrinos in  neutrino beam  experiments.   
\end{abstract}


\section{Introduction}

Despite its huge successes, the Standard Model (SM) of particle physics has several drawbacks which require one to conceive some Beyond the SM (SM) physics. Its  Achilles' heel is probably the leptonic sector, though, as neutrino masses are forbidden in the SM, yet, experiments have verified that neutrino flavours 
oscillate which  in turn implies that neutrinos have finite masses.  Neutrinos are strictly massless in the SM essentially due to two reasons: 
$(i)$ the absence of their right-handed eigenstates; $(ii)$ an exact global Baryon minus Lepton $(B-L)$ number conservation. However, a
  modification of the SM, based on the gauge group $SU(3)_C \times SU(2)_L
\times U(1)_Y \times U(1)_{B-L}$, nicknamed the $B-L$ extension of the SM (BLSM), wherein the additional Abelian group is elevated to be a local symmetry, 
can account for light neutrino masses through  an  Inverse Seesaw (IS) mechanism \cite{Khalil:2006yi,Khalil:2010iu}.
In such a construct, the aforementioned right-handed neutrinos would acquire Majorana masses at the $B-L$ symmetry breaking scale, but they are not allowed to do so
by the discussed $B-L$ gauge symmetry and another pair of SM gauge singlet
fermions with tiny masses, of ${\cal O}(1$ keV), must be introduced. 
Therefore, such a small 
scale can be considered as a slight breaking of the underlying gauge symmetry, hence,  according to 't Hooft criteria, its dynamics  becomes natural.

One of these two singlet fermions couples to right-handed neutrinos and is involved in generating the light neutrino masses. The other singlet (usually called inert or sterile neutrino) is completely decoupled and interacts
only through the $B-L$ gauge boson, a $Z'$, ensuing from the spontaneous breaking of the additional $U(1)_{B-L}$ group \cite{Elsayed:2011de}, so that  
 it may account for warm Dark Matter (DM) \cite{El-Zant:2013nta} (see also Ref. \cite{Basso:2012gz}), the lack of a viable candidate  for it being another significant flaw of the SM. This construct,  
 BLSM-IS for short, predicts several testable signals at the Large Hadron Colider (LHC) through some of the new particles that it embeds: the $Z^\prime$ (neutral gauge boson) associated with  $U(1)_{B-L}$, an extra Higgs boson ($h'$, in fact, an additional (pseudo)scalar singlet state is introduced to break the gauge group $U(1)_{B-L}$
spontaneously) and heavy neutrinos ($\nu_h$, which are required to cancel the associated new gauge anomalies and are thus necessary for the consistency of the whole model). 

Ref. \cite{Khalil:2015naa} reviewed the LHC potential to access the BLSM-IS, including its Supersymmetric extension \cite{Khalil:2012gs,Khalil:2013in,Moretti:2019ulc}, when the $Z'$ mass is of order TeV and such a state is relatively strongly coupled to SM states.  
In this paper, we aim instead at considering the case of a very light $Z'$, of MeV scale, very mildly coupled to SM objects, specifically, whether it can be responsible, together with 
the aforementioned sterile neutrinos, of data anomalies that have emerged from the  E821 experiment at BNL and the Muon $g-2$ one at FNAL as well as the MiniBooNE (MB) collaboration also at FNAL. In fact, the former two hinted at statistically significant deviations from the SM predictions of the anomalous magnetic moment of the muon, $(g-2)_\mu$ for short, which could be explained by a very light $Z'$ state, while the latter one was taken as a sign of  the possible existence of sterile neutrinos. 

The plan of the paper is as follows. In the next section, we describe the BLSM-IS. In Sect.~III, we discuss both direct and indirect experimental constraints on 
light $Z'$ and sterile neutrino states. We then move on to present our results for $(g-2)_\mu$.  After this, we discuss our explanation for the MB excess. Finally, in the last section, we present our summary.

\section{The model}

To start with, in the BLSM-IS that we consider here, we assume that the SM singlet scalar $\chi$, which spontaneously breaks $U(1)_{B-L}$,
has $B-L$ charge $= -1$. Also, the three pairs of SM singlet fermions, $S_{1,2}$ with $ B-L$ charge $= \mp 2$, respectively, are introduced (see tab.~I, wherein $l_{L,R}$ refer to leptons, $Q_L,u_R,d_R$  identify quarks and $\phi$ is the Higgs state of the SM).  
\begin{table}[!t]
\label{tab:1}
\addtolength{\tabcolsep}{8pt}
\begin{tabular}{ |c|c|c|c|c|c|c|c|c|c|c|}\hline 
Particle &$ l_L$ & $l_R $&$Q_L$&$u_R$&$d_R$&$\nu_R$&$\phi$&$\chi$&$S_1$&$S_2$ \\\hline
$B-L$ charge & -1& -1 &$\frac{1}{3}$&$\frac{1}{3}$&$\frac{1}{3}$&-1&0&-1&-2&+2\\\hline
\end{tabular}
\caption{$B-L$ quantum numbers for the BLSM-IS model.}
\end{table}
Limited to the leptonic sector, the BLSM-IS  Lagrangian is given by
\begin{equation}\label{eq:1}
\begin{split}
\mathcal{L}_{(B-L)} &= -\frac{1}{4}F^\prime_{\mu\nu}F^{\prime\mu\nu} +i\bar{l}_LD_\mu\gamma^\mu l_L+i\bar{l}_R D_\mu\gamma^\mu l_R + i \bar{S_1}_R D_\mu \gamma^\mu S_1+  i \bar{S_2}_R D_\mu \gamma^\mu S_2+
i\bar{\nu_R}D_\mu\gamma^\mu\nu_R\\
& +(D^\mu\phi)^\dagger (D_\mu\phi)+ (D^\mu\chi)^\dagger (D_\mu\chi)
-V(\phi,\chi)-\left(\lambda_l\bar{l}_L\phi l_R+ \lambda_\nu \bar{l}_L\tilde{\phi} \nu_R+\lambda_s \bar{\nu}_R \chi S_2\right)+ h.c. ,
\end{split}
\end{equation}
with $\tilde{\phi}=i\sigma^2\phi^\ast$. Using the unitary gauge parameterisation, the kinetic terms become
\begin{equation}
\left(D^\mu\phi\right)^\dagger\left(D^\mu \phi\right) =\frac{1}{2}\partial^\mu h\partial_\mu h+\frac{1}{8}\left(\phi+\upsilon \right)^2\left[g^2|W^\mu_1-iW^\mu_2|^2+\left(gW^\mu_3-g_1B^\mu-\tilde{g}B^{\prime\mu} \right)^2\right]
\end{equation}
and
\begin{equation}
\left(D^\mu\chi\right)^\dagger\left(D^\mu \chi\right) =\frac{1}{2}\partial^\mu h^\prime\partial_\mu h^\prime+\frac{1}{2}\left(h^\prime+\upsilon^\prime \right)^2\left(2g_{(B-L)}B^{\prime\mu}\right),
\end{equation}
where $g_{(B-L)}$ is the coupling strength of the new $Z'$ boson, 
$\tilde{g}$ is the gauge kinetic mixing parameter and $\upsilon,\upsilon^\prime$ are the SM and $U(1)_{B-L}$ vacuum expectation values. The mass eigenstates of the gauge boson fields are linear combinations of $B^\mu,W_3^\mu$ and $B^{\prime\mu}$. The explicit expression for the mass mixing matrix is
\begin{equation}
\left(\begin{array}{c}
B^\mu\\W_3^\mu\\B^{\prime\mu}
\end{array}\right) = \left(\begin{array}{ccc}
\cos\theta_W &-\sin\theta_W\cos\theta^\prime&\sin\theta_W\sin\theta^\prime\\
\sin\theta_W&\cos\theta_W\cos\theta^\prime&-\cos\theta_W\sin\theta^\prime\\
0&\sin\theta^\prime&\cos\theta^\prime
\end{array}\right)\left(\begin{array}{c}
A^\mu\\Z^\mu\\Z^{\prime\mu}
\end{array}\right)\,,
\end{equation}  
with $\theta_W$  the weak mixing angle while $\frac{-\pi}{4}\le\theta^\prime\le\frac{-\pi}{4}$ such that
\begin{equation}
\tan 2\theta^\prime=\frac{2\tilde{g}\sqrt{g^2+g_1^2}}{\tilde{g}+16\left(\frac{\upsilon^\prime}{\upsilon} \right)^2g^2_{(B-L)}-g^2-g_1^2}\,.
\end{equation}
The neutral gauge boson masses are determined by fixing the values of the new parameters as
\begin{equation}
M_{Z,Z^\prime}=\frac{\upsilon\sqrt{g^2+g_1^2}}{2}\left[\frac{1}{2} \left(\frac{\tilde{g}^2+16\left(\frac{\upsilon^\prime}{\upsilon} \right)^2g^2_{(B-L)}}{g^2+g^2_1}+1 \right) \mp \frac{\tilde{g}}{\sin 2\theta^\prime\sqrt{g^2+g^2_1}}\right]^\frac{1}{2}\,.
\end{equation}

In the  BLSM-IS model, the Majorana neutrino Yukawa interaction induces the  masses onto the SM neutrinos after $U(1)_{B-L}$ symmetry breaking via the Lagrangian terms
\begin{equation}
{\mathcal{L^\nu}} = m_D\bar{\nu_L}\nu_R + m_N \bar{\nu}_R S_2 + h.c\,,
\end{equation}
with a Dirac mass $m_D=\frac{1}{\sqrt{2}} \lambda_\nu \upsilon$ and a Majorana mass $m_N = \frac{1}{\sqrt{2}} \lambda_S \upsilon^\prime$. The $9\times 9$ neutrino mass matrix can be written as 
\begin{equation}
\mathcal{M}_\nu =\left(\begin{array}{ccc}
0&m_D&0\\
m_D^T&0&m_N\\
0&m_N^T&\mu_s\\
\end{array}\right)\,.
\end{equation}
In order to  avoid a possible large mass term $m S_1 S_2$ in the Lagrangian, that would spoil the IS structure, one assumes a $Z_2$ symmetry under which  $\nu_R, \chi, S_2$  and the SM particles are even while $S_1$ is an odd particle. The neutrino mass matrix $\mathcal{M}_\nu$ can be diagonalised by the  matrix $V$ as\footnote{Assuming there are no complex Majorana phases and the Lagrangian parameters are real.}
\begin{equation}
V^T \mathcal{M}_\nu V = \mathcal{M}^{\text{diag}}_\nu \, ,
\end{equation}
with $V$ a $9\times 9$ matrix defined as \cite{Abdelalim:2014cxa}
\begin{equation}
V =\left(\begin{array}{cc}
V_{3\times 3}&V_{3\times 6}\\
V_{6\times 3}&V_{6\times 6}
\end{array}\right)\,.
\end{equation}
The upper $3\times 3$ block are the parameters for  the effective Pontecorvo-Maki-Nakagawa-Sakata (PMNS) matrix with its elements given by 
\begin{equation}
V_{3\times 3 } = \left( 1- \frac{1}{2}\theta^T \theta\right) U_{\text{PMNS}},
\end{equation}
in terms of the actual one.
The off-diagonal blocks of the $V$ matrix are defined via 
\begin{equation}
V_{3\times 6 } = \left( 0_{3\times 3} ,\theta\right) V_{6\times 6}\,,
\end{equation}
with $\theta\sim m_D m_N^{-1}$.  The matrix $V_{6\times 6}$  diagonalises the  right-handed Majorana neutrinos and $S_2$.
The diagonalisation of the entire  neutrino mass matrix  leads to the following light and heavy neutrino masses:
\begin{equation}
\begin{split}
m_{\text{light}} &\sim m_D m_N^{-1}\mu_s (m_N^T)^{-1}m_D^T\,,\\
m_{\text{heavy}} &\sim \left(m_N^2+m_D^2 \right)^{\frac{1}{2}} \,,
\end{split}
\end{equation}
with the latter being pair degenerate. With this structure,  the light neutrinos can be of order eV, as required by flavour oscillation experiments,  and, with a small $\mu_s$ value, the ensuing Yukawa coupling is no longer restricted to be very small, indeed,
it can be   of order one. Moreover,  the mixing between light and heavy neutrinos is constrained from lepton flavour violation measurements  to be  of order $\mathcal{O}(0.01)$ as discussed in \cite{Antusch:2015mia,Antusch:2020vul,Alonso:2012ji} and references therein.

The tree level coupling of the $Z^\prime$ with charged and neutral fermions is expressed as
\begin{eqnarray}\label{eq:9}
g_{(Z^\prime\bar{l}_i l_j)} &=&  \frac{i\delta_{ij}}{2}\left[2\left(g_{(B-L)}\cos\theta^\prime+\tilde{g}\sin\theta_W\sin\theta^\prime \right)+g_1\sin\theta_W\sin\theta^\prime-g_2\cos\theta_W\sin\theta^\prime \right]\gamma^\mu \left(\frac{1-\gamma_5}{2} \right) \nonumber\\
&+&i\delta_{ij}\left[g_{(B-L)}\cos\theta^\prime+(g_1 + \tilde{g})\sin\theta_W\sin\theta^\prime \right]\gamma^\mu \left(\frac{1+\gamma_5}{2} \right),
\end{eqnarray}
while the ones with active and sterile (light and heavy) neutrinos are given by
\begin{equation}\label{eq:2}
\begin{split}
g_{(Z^\prime,\nu_i,\nu_j)} &= \frac{i}{2} \left\lbrace  \left[2g_{(B-L)}\cos\theta^\prime + (2\tilde{g} +g_1)\sin\theta_W\sin\theta^\prime+g_2\cos\theta_W\sin\theta^\prime  \right] \sum_{a=1}^3 V_{ja}^* V_{ia}\right.  \\
&\left. -\left[2g_{(B-L)}\cos\theta^\prime +2 \tilde{g} \sin\theta_W\sin\theta^\prime  \right]\sum_{a=1}^3 V_{j3+a}^* V_{i3+a}\right\rbrace \, \gamma^\mu \left( \frac{1-\gamma_5}{2}\right)+\\
&+(-\frac{i}{2}) \left\lbrace  \left[2g_{(B-L)}\cos\theta^\prime + (2\tilde{g} +g_1)\sin\theta_W\sin\theta^\prime+g_2\cos\theta_W\sin\theta^\prime  \right] \sum_{a=1}^3 V_{ia}^* V_{ja}\right.  \\
&\left. -\left[2g_{(B-L)}\cos\theta^\prime +2 \tilde{g} \sin\theta_W\sin\theta^\prime  \right]\sum_{a=1}^3 V_{i3+a}^* V_{j3+a}\right\rbrace \, \gamma^\mu \left( \frac{1+\gamma_5}{2}\right) ,
\end{split}
\end{equation}
with $\theta^\prime$ constrained from LEP experiment to be $\sim 3\times 10^{-3}$ \cite{Abreu:267165}.

\section{Direct and Indirect constraints on light $Z^\prime$ and sterile neutrinos}
\begin{figure}[!t]
\centering
\includegraphics[scale=0.3]{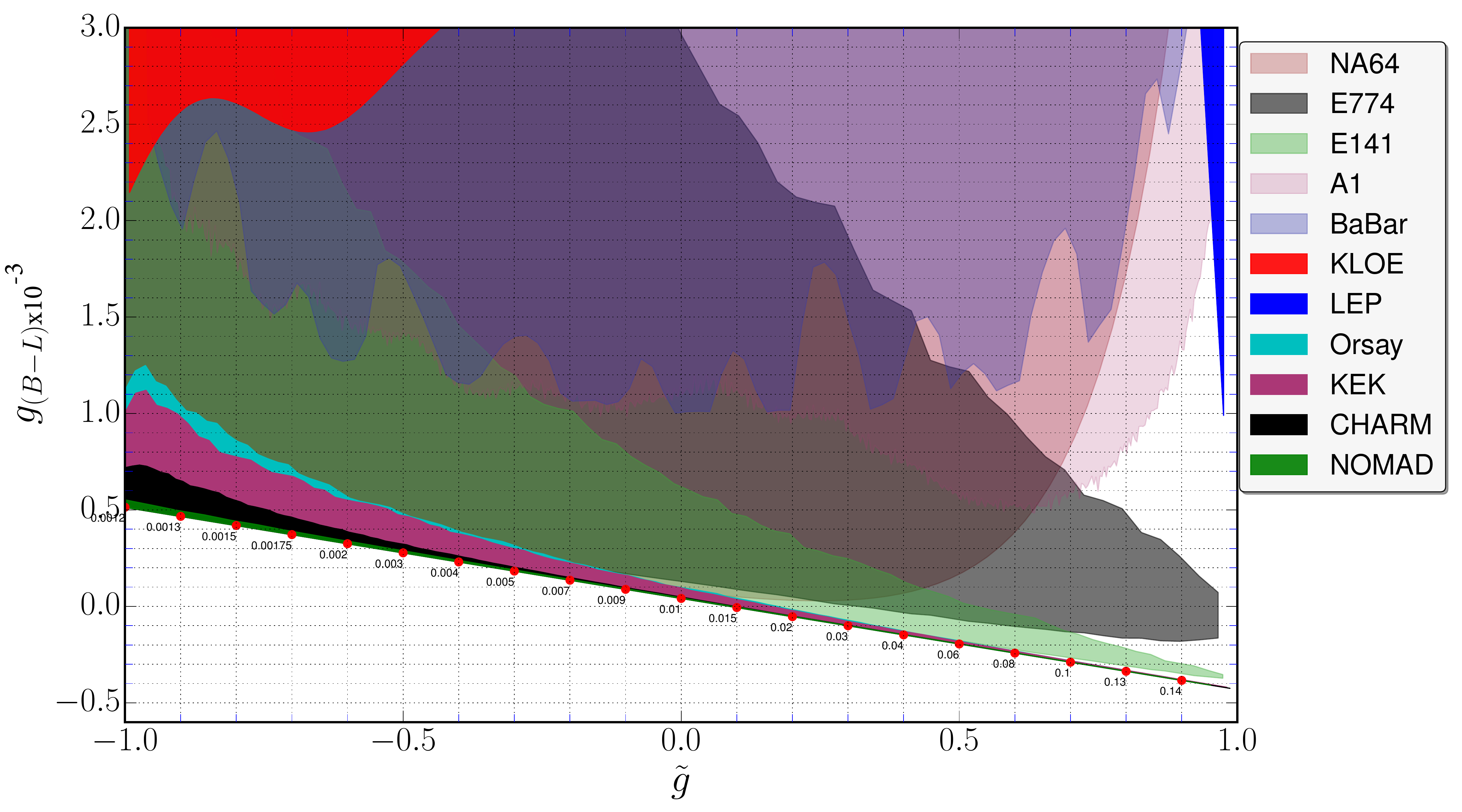}
\caption{Bounds on the plane $(\tilde{g}, g_{(B-L)})$ for different $Z^\prime$ mass values in the BLSM-IS for fixed $\theta^\prime=3\times 10^{-3}$. The allowed region is the non-shaded area with allowed values of $M_{Z^\prime}$ less than each corresponding value (in GeV) on the red dots.}
\label{fig:1}
\end{figure}

In this section we discuss the direct and the indirect constraints for low mass $Z^\prime$ and sterile neutrinos. In fig. \ref{fig:1} we show the most severe constraints on the light $Z^\prime$ mass as a function of the $Z^\prime$ gauge coupling $g_{(B-L)}$ and  the gauge kinetic mixing parameter $\tilde{g}$ from existing low energy experiments.   
To recast these bounds on those applicable to our model we used the method of Ref.~\cite{Ilten:2018crw} and produced fig. \ref{fig:1} by using the code advertised in the same paper.  The key here is that low energy experiments setting bounds on the low mass photon (a dark photon, $A'$) considered therein also set limits on a light dark $Z^\prime$, so long that one accounts for the gauge kinetic mixing and axial coupling.
Recasting a dark photon search that used the final state $F$ in constraints onto our model can be done, for each $Z^\prime$ mass, by  equating  the upper limit total cross section of dark photon models to the $Z'$ one in our model as  follows:
\begin{equation}
\sigma_{Z^\prime}\times {\rm BR}(Z^\prime\to F) \times\epsilon_{Z^\prime} = \sigma_{A^\prime}\times {\rm BR}(A^\prime\to F) \times\epsilon_{A^\prime},
\end{equation} 
with $\sigma_{Z^\prime/A^\prime}$ being the production cross section and BR$(Z^\prime/A^\prime\to F)$  the Branching Ratio (BR) of the light gauge boson into the final state $F$ while $\epsilon_{Z'/A'}$ is the detector efficiency. Therefore, one can see that, in order to recast the aforementioned experimental limits in terms of our model parameters, we only need the ratios $\sigma_{Z^\prime}/\sigma_{A^\prime}$, BR$(Z^\prime\to F)/{\rm BR}(A^\prime\to F)$ and $\epsilon_{Z^\prime} /\epsilon_{A^\prime} $. 

In the following, we are  going to discuss how these ratios can be obtained for each experiment.
\begin{itemize}
\item The BaBar detector at the PEP-II B-factory \cite{BaBar:2017tiz} has collected 53 fb$^{-1}$ of $e^-e^+$ collisions looking for events with a single high-energy photon and large
missing (transverse) momentum or energy which is consistent with the process $e^- e^+\to\gamma X$ and $X\to\text{invisible}$, with $X$ being a light gauge boson with spin equal to 1. Further, in \cite{BaBar:2014zli}, the BaBar experiment searched  for a single high energy photon plus a dilepton final state,   $e^- e^+\to\gamma X$ and $X\to\bar{l}l$, with $l=e,\mu$. In both searches  no statistically significant deviations from the SM
predictions have been observed and a $90\%$ Confidence Level (CL) upper limit on the light gauge boson coupling to leptons in the mass range of $0.02 -10.2$ GeV has been set. Recasting this limit onto our model we obtain 
\begin{equation}\label{eq:12}
 \frac{g^2_{Z^\prime}\times{\rm  BR}(Z^\prime\to ll)}{g^2_X\times {\rm BR}(X\to ll)} =1\,,
\end{equation}   
with $g_{Z^\prime}$ being the $Z^\prime$ coupling to charged and neutral leptons, eqs. (\ref{eq:1}) and (\ref{eq:2}), and $g_X$ being  the measured gauge boson coupling to charged and neutral leptons.
\item The A1 Collaboration at the Mainz Microtron (MAMI) \cite{Merkel:2014avp} searched for the signal of a new light $U(1)$ gauge boson in electron-positron pair production. Since no deviation from the SM value for the corresponding cross section has been observed, A1 set a limit on the light gauge boson coupling over the mass range $40-300$ MeV.  To recast this limit on our model parameters, we have again made use of  eq. (\ref{eq:12}).
\item Electron beam dump experiments (like E141, E774 and those at KEK and Orsay) also have sensitivity to a new light gauge boson. An overview of the different electron beam dump experiments and their properties is given in \cite{Andreas:2012mt}. For the SLAC E141 experiment \cite{Riordan:1987aw}, an upper limit is set for  neutral particles with masses
in the range $1-15$ MeV following the non-observation of any excess above the SM  bremsstrahlung rate for events of the type $e+ N \to e+N+X$. From the Fermilab E774 experiment \cite{Bross:1989mp}, an upper limit for neutral particles which decay into electron-positron pairs was set. In the electron beam dump experiment at KEK \cite{Konaka:1986cb}, no signal was observed in their search for axion-like particles. The electron beam dump experiment in Orsay \cite{Davier:1989wz} also found no positive signal when looking for light Higgs bosons decaying into electron-positron pairs. Combining and reinterpreting these last three experiments, one is able to exclude a light boson over the mass range $1.2-52$ MeV.
\item Proton beam dump experiments, like NOMAD \cite{NOMAD:2001eyx} and CHARM \cite{CHARM:1985anb}, also  found no positive signal while looking for axion like-particles decaying to leptonic pairs, following which the $0.1-20$ MeV mass range is also precluded to a dark photon or $Z'$.
\item The NA64 experiment at the CERN SPS \cite{NA64:2016oww} found no deviation from the SM expectation while  looking for dark photons in the process $e^- N\to e^- N A^\prime$. Hence, a new limit has been set on the $A'$ (dark photon) mixing and the absence of  invisible $A^\prime$ decays excluded the mass range $M_{A^\prime} \le 100$ MeV. 
\item The DELPHI experiment at LEP2 \cite{DELPHI:2008uka} analysed single photon events in looking for extra dimension gravitons. As in \cite{Freytsis:2009bh}, since the measured single-photon cross sections are in agreement with the expectations from the SM,  an upper limit on the coupling and mass of the dark candidate was set, the latter being above $10$ GeV. 
\end{itemize} 

{Before moving on to study the relevant experimental observables, we should mention that 
we have used SPheno \cite{Porod:2011nf,Porod:2003um} to generate the model spectrum as well as HiggsBounds and HiggsSignals \cite{Bechtle:2008jh,Bechtle:2011sb,Bechtle:2012lvg,Bechtle:2013wla,Bechtle:2015pma} to check the constraints on the Higgs  sector of it. Also, we have used FlavourKit \cite{Porod:2014xia} to check lepton flavour violation constraints.
}

\section{The muon anomalous magnetic moment}
The Lande $g$ factor for muons, and its deviation from the tree
level value of 2, represents one of the most precisely measured quantities in the SM. Therefore, it is also an excellent probe for new physics. Currently, there exists a long standing and statistically significant discrepancy between
its measurement and the theoretically predicted
value \cite{Muong-2:2021ojo,Davier:2019can,Davier:2017zfy,Davier:2010nc}\footnote{A recent lattice calculation \cite{Borsanyi:2020mff} is suggesting a somewhat different value for the $(g -2)_\mu$ value, so that, if such an estimate
is correct, the deviation between measurement and theory is smaller than $4.2\sigma$.}:
\begin{equation}
\Delta a_\mu = \Delta a^{\rm ex}_\mu-\Delta a^{\rm th}_\mu= (2.51 \pm 0.59)\times 10^{-9}. 
\end{equation}
In this section, we focus on a light $Z^\prime$ as a means to solve the current muon anomalous magnetic moment anomaly.  
Following the general formula in \cite{Moore:1984eg}, the interaction Lagrangian of a $Z^\prime$ with muons can be rewritten as 
\begin{equation}
\mathcal{L}_{\rm int} = \bar{\mu}\gamma^\mu\left(C_V+C_A\gamma^5 \right)\mu Z^\prime\,,
\end{equation}
where $C_V$ and $C_A$ are the vector and axial couplings introduced in eq. (\ref{eq:9}). 
\begin{figure}[!h]
\centering
\includegraphics[scale=0.9]{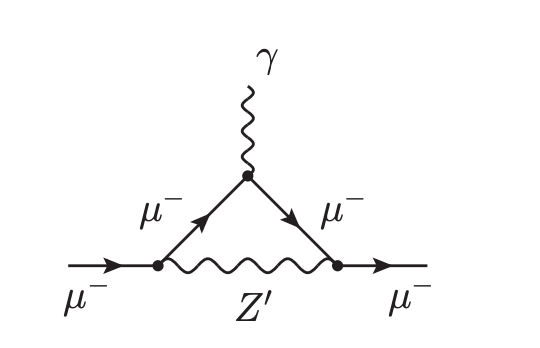}
\caption{Feynman diagram for the $Z^\prime$ contribution to the muon anomalous magnetic moment.}
\label{fig:2}
\end{figure}
The $Z^\prime$ modifies the muon magnetic moment via the one loop diagram in fig. \ref{fig:2}. The $Z^\prime$ contribution can be obtained as \cite{Bodas:2021fsy}
\begin{equation}
\Delta a_\mu = \frac{m^2_\mu}{4\pi^2m^2_{Z^\prime}}\left[C^2_V\int^1_0\frac{x^2(1-x)}{1-x+x^2\frac{m^2_\mu}{m^2_{Z^\prime}}}dx -C^2_A \int^1_0\frac{x(1-x)(4-x)+2x^3\frac{m^2_\mu}{m^2_{Z^\prime}}}{1-x+x^2\frac{m^2_\mu}{m^2_{Z^\prime}}}dx  \right]\,,
\end{equation}
with $x$ being the Feynman parameter. For a low mass $Z^\prime$, its  contribution to the muon anomalous magnetic moment is
\begin{equation}
\Delta a_\mu \simeq \frac{(m^2_{Z^\prime} C^2_V -2 m^2_\mu C^2_A)}{8\pi^2m^2_{Z^\prime}}.
\end{equation}

\begin{figure}
\centering
\includegraphics[scale=0.5]{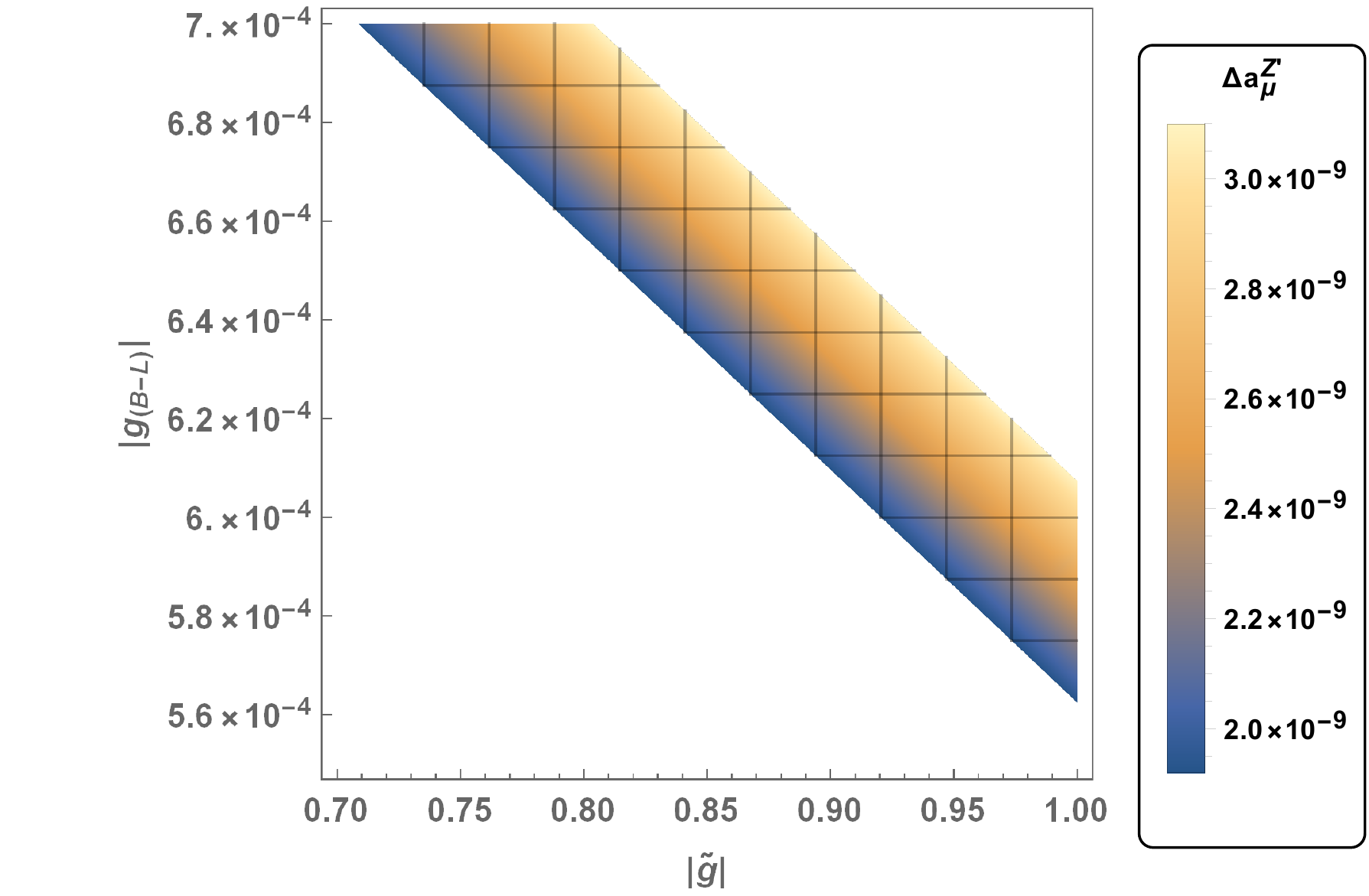}
\caption{The $\Delta a_\mu^{Z'}$  dependence on the two coupling $\tilde{g}$ and $g_{(B-L)}$.}
\label{fig:3}
\end{figure}
Fig.~\ref{fig:3}  shows the $\Delta a_\mu^{Z'}$  dependence on the two coupling $\tilde{g}$ and $g_{(B-L)}$. The density plot is confined between the upper and lower experimental values $((2.51 \pm 0.59)\times 10^{-9})$, respectively, of $\Delta a_\mu^{Z'}$ within $1\sigma$ CL. The  plot  represents the allowed region of the model parameters that satisfies the experimental data on $\Delta a_\mu^{Z'}$. Here, $|\tilde{g}|$ is taken from $10^{-4}$ to $10^{-1}$ with the allowed range of $|g_{(B-L)}|$ being from $(9.93 -10.79) \times 10^{-4}$ to $(9.87 -10.86) \times 10^{-4}$, respectively.  The corresponding range of vector and axial couplings for $M_{Z^\prime}=30$ MeV and $|\tilde{g}|=10^{-4}$ are $|C_V| = (10.14-10.59)\times 10^{-4}$ and  $|C_A| \approx 1.88\times 10^{-4}$ whereas for $|\tilde{g}|=10^{-1}$ they are $|C_V| = (9.67-11.06)\times 10^{-4}$ and  $|C_A| = 1.88\times 10^{-4}$. This  viable region of model parameters is also compliant with the constraints given in \cite{Bodas:2021fsy}, which included the following ones. 
\begin{enumerate}
\item Cosmological and astrophysical bounds: Big Bang Nucleosynthesis (BBN) \cite{ParticleDataGroup:2018ovx} as well as Cosmic Microwave Background (CMB) from ``Planck 2018'' \cite{ParticleDataGroup:2018ovx} in addition to the astrophysical experiments (e.g SN1987A) studied by \cite{Croon:2020lrf}.
\item Neutrino scattering bounds: several neutrino scattering experiments results on couplings to muons and muon neutrinos,
the most stringent ones of these being from Borexino \cite{Bellini:2011rx} and CHARM-II \cite{CHARM-II:1993phx}.
\end{enumerate}

Observation of energy loss in supernovae due to $Z′ − μ$ interactions set constraints on the $B-L$ model parameters \cite{Croon:2020lrf,Knapen:2017xzo,Chang:2016ntp,Rrapaj:2015wgs}. A $Z'$ mass up to 100 MeV is constrained in the $(M_{Z'},g_{(B-L)})$ plane \cite{Croon:2020lrf}. Both cosmological (BBN) and astrophysical (SN1987A) limits are model dependent. For instance, the chameleon effect due to the environmental matter density and late reheating can weaken the SN1987A \cite{Nelson:2008tn} and BBN \cite{Dev:2019hho} limits, respectively. In the study of the astrophysical limit, a pure $Z'$ model has been considered \cite{Croon:2020lrf}, but in our model we have an extended scalar sector. In the presence of new scalar states, the limits on $Z'$ change dramatically and can be avoided when a neutral state couples to a dark matter particle \cite{Zhang:2014wra}, as is the case in our model.

A model-independent fit to all such experimental data (thus including $(g-2)_\mu$) reveals the following parameter values as viable.
	\begin{enumerate}
		\item A light $Z'$ in the mass range $16 \; {\rm MeV} \lesssim M_{Z'} \lesssim 38 \; {\rm MeV}$.
		\item An axial coupling of the $Z'$ to electrons larger than the vector one: $|C_{Ae}| \sim [1-3.2] \times 10^{-4} > \, |C_{Ve}| \lesssim 7.7\times 10^{-5}$.
		\item A large vector coupling to muons, $5 \times 10^{-4} < |C_{V\mu}| \lesssim 0.05$, and an axial coupling $C_{A\mu}$ that is smaller by at least a factor of a few.
		\item Tiny $Z'$ couplings to neutrinos: $|C_{\nu_e,\, \nu_\mu}| \lesssim 10^{-5}$.
	\end{enumerate}

We now move on to study the MB anomaly and its theoretical implications.

\section{MiniBoone}

\begin{figure}
\centering
\includegraphics[scale=0.25]{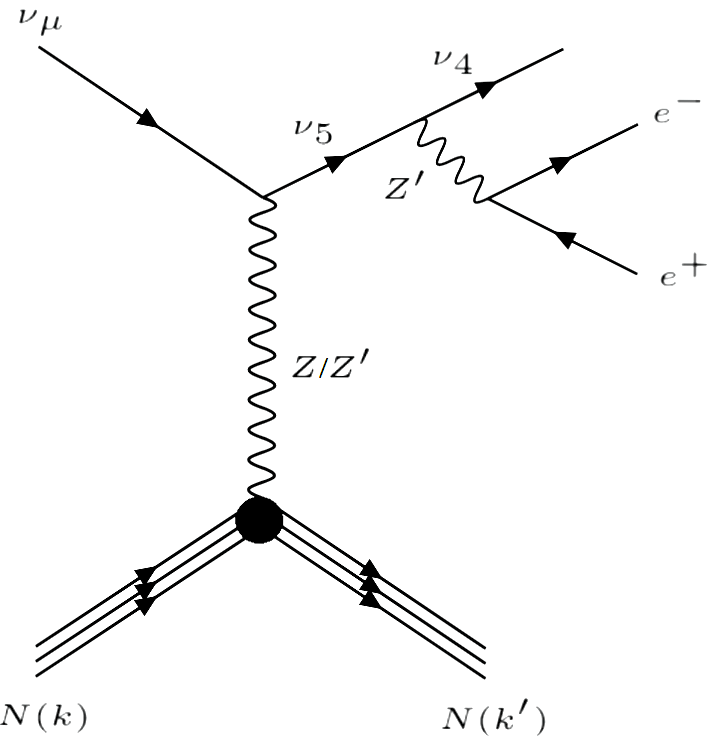}
\caption{Feynman diagram of the scattering process in the BLSM-IS which leads to the excess in MB.}
\label{fig:4}
\end{figure}

In this section, we will study the anomaly registered by the MB experiment, wherein the  beam  primarily 
consists of $\nu_{\mu}$'s produced via pion decay. The relevant process  
leading to an electron excess is $\nu_\mu N(k) \rightarrow N(k') \nu_4 
e^{+} e^-{-}$, as 
shown in fig.~\ref{fig:4}, where $\nu_{4,5}$ are sterile neutrinos, with $m_{\nu_5}>m_{\nu_4}$. ({{Previous work explaining the results in Ref.~\cite{MiniBooNE:2020pnu} using sterile neutrinos can be found in \cite{Fischer:2019fbw,Jordan:2018qiy,Bertuzzo:2018itn,Ballett:2018ynz,Arguelles:2018mtc,Chang:2021myh,Abdallah:2020vgg,Abdallah:2020biq,Blanco:2019vyp,Dentler:2019dhz,deGouvea:2019qre,Brdar:2021cgb,Doring:2018cob,Abdullahi:2020nyr,Dutta:2020scq,Datta:2020auq,Giunti:2019sag}}.) The process is mediated by the new 
$Z'$ producing a collimated $e^+ e^-$ pair (producing the visible light that makes up the signal) through the decay of $\nu_5$ into $\nu_4$, which then makes the cross section proportional to $|
U_{\nu_5 \nu_4}|^2$. 
By calculating the liftime of $\nu_4$, it is found to be 1.76 second. The corresponding decay length is $5.3 \times 10^8$ m which is way greater than the MB dimention of 12.2 m. Thus, $\nu_4$ will decay away outside the detector. So that there is no additional EM deposition in the detector due to the $\nu_4$ decay. 
The form factor of the coupling of the  $Z'$ 
with nucleons $N$ is
\begin{eqnarray} \nonumber
\langle N(k')|J^{\mu}_{Z'}| N(k)\rangle &=& g_{B} \bar{u}(k')  \Gamma^{\mu}_{Z'} (k'-k)u(k),
\end{eqnarray}
where $k$ and $k'$ are the initial and final nucleon momenta  whereas   
\begin{eqnarray}\label{eq:17}
\Gamma^{\mu}_{Z'}(q) = \gamma^{\mu} F^{1}_{V}(q^2)  + \dfrac{i}{\,2 \, m_{N}}\sigma^{\mu \nu} q_{\nu} \, F^{2}_{V}(q^2).
\end{eqnarray}
The isoscalar form factors  $F^{1}_{V}(q^2)$ and $ F^{2}_{V}(q^2)$ for the nucleon are given by~\cite{Hill:2010yb}
\begin{equation}
\dfrac{F^{1}_{V}(q^2)}{F_{D}(q^2)}  =  1 - \dfrac{q^2 (a_{p}+ a_{n})}{4m^{2}_{N}- q^2},~~~~~\frac{F^{2}_{V}(q^2)}{F_{D}(q^2)} =  \dfrac{4m^{2}_{N}(a_p + a_n)}{4m^{2}_{N}- q^2 },
\end{equation}
where $m_N=0.938$~GeV, $F_{D}(q^2) = (1-q^{2}/0.71~\rm{GeV^{2}})^{-2}$ with $a_p \approx 1.79$ and $a_n  \approx -1.91$ being coefficients related to the  magnetic moments of the proton and neutron, respectively.

The total differential cross section has two components, an incoherent and a coherent one, which we will both consider. The total differential cross section, for the target in  MB,  i.e., CH$_2$, is given by
\begin{eqnarray}
\left(\dfrac{d\sigma}{dE_{h'}}\right)_{{\rm CH_{2}}} =\underbrace{14 \times \left(\dfrac{d\sigma}{dE_{h'}}\right)}_{\textrm{\footnotesize{{incoherent}}}}+ \underbrace{ 144 \times {\rm exp}(2b(k'-k)^{2})\left(\dfrac{d\sigma}{dE_{h'}}\right)}_{\textrm{\footnotesize{{coherent}}}}.
\label{tot_xsec}
\end{eqnarray}
The incoherent contribution from the single nucleon cross section is multiplied by the total number of the nucleons present in CH$_2$, {i.e.}, 14. However, the entire carbon nucleus contributes to the coherent process weighted by the exponential factor exp$(2b(k'-k)^{2})$~\cite{Hill:2009ek}, where $b$ is a numerical parameter, which for C$^{12}$ has been found  to be $25$~GeV$^{-2}$~\cite{Freedman:1973yd, Hill:2009ek}. The coherent process decreases as $q^2 = (k'-k)^2$ increases, where $q^2$ is negative.


The number of events is given by \cite{Vergani:2021tgc}
\begin{equation}
{\rm N}_{\rm{events}} \! =\! \eta\!\! \int\!\! dE_{\nu} dE_{\nu_5}\dfrac{d\Phi^{\rm{\nu}}}{dE_{\nu}} \!\,  \dfrac{d\sigma} {dE_{\nu_5}}\! \times\! {\rm BR}(\nu_5 \!\rightarrow\! \nu_4 Z^\prime)\! \times\! {\rm BR}(Z^\prime \!\rightarrow\! e^- e^+)\times n,
\end{equation}
with $E_{h'}\in [E_{h'},E_{h'}+\Delta E_{h'}]$ and where $\Phi^{\rm{\nu}}$ is the incoming muon neutrino flux. Here, $n$  is the number of nuclei in the fiducial volume of the detector. In the case of MB, the target is 818 tons of mineral oil (CH$_2$) with atomic mass  14 \cite{MiniBooNE:2020pnu}, as mentioned, so that $n=3.5174 \times 10^{31}$.  Furthermore, $\eta=0.2$ contains all the detector related  information like efficiencies, Protons-on-Target (POT), etc. 
The latest data set  for the neutrino mode, corresponding to $18.75 \times 10^{20}$ POT, as detailed in \cite{MiniBooNE:2020pnu,Vergani:2021tgc}, has been used in our fit.
Finally, for these  values, the calculated  lifetimes of the $\nu_5$ and $Z'$ states in their rest frame are $10^{-17}$~s and $1.8\times 10^{-12}$~s, respectively.

The value of $E_{\nu_5}$ is related to the visible energy, $E_{\rm vis} =E_{e^+}+E_{e^-} $, as follows
\begin{equation}
E_{\nu_5}=\frac{E_{Z^\prime} \left(-M_{\nu_4}^2+M_{\nu_5}^2+M_{Z^\prime}^2\right)-\sqrt{\left(E_{Z^\prime}^2-M_{Z^\prime}^2\right) \left(M_{\nu_4}^4-2 M_{\nu_4}^2 \left(M_{\nu_5}^2+M_{Z^\prime}^2\right)+\left(M_{\nu_5}^2-M_{Z^\prime}^2\right)^2\right)}}{2 M_{Z^\prime}^2}.
\end{equation}
Furthermore, the Mandelstam variables in terms of the neutrino(lepton) energy $E_\nu$($E_l$) are
\bea
\label{Manvar}
s &= & M^2 + 2 M E_\nu,  \nonumber\\ t& =& 2 M(E_l -E_\nu),  \nonumber\\  s-u &=& 4 M E_\nu + t -m^2_l.
\eea
 Then, $t$ and $E_l$ lie in the intervals 
 \bea
 \label{tint}
  m^2_l- 2 E_\nu^{\rm cm} \left( E_l^{\rm cm} + p_l^{\rm cm} \right)   \leq  t \leq  
  m^2_l- 2 E_\nu^{\rm cm} \left( E_l^{\rm cm} - p_l^{\rm cm}\right) ,
 \eea
 \bea
 E_\nu + \frac{m_l^2-2E_\nu^{\rm cm} (E_l^{\rm cm} + p_l^{\rm cm})}{2M} \leq  E_l \leq E_\nu + \frac{m_l^2-2E_\nu^{\rm cm} (E_l^{\rm cm} - p_l^{\rm cm})}{2M},
 \eea
where the energy  and momentum of the  neutrino and lepton  in the center of mass (cm) system are
\bea
E_\nu^{\rm cm} &=& \frac{(s-M^2)}{2 \sqrt{s}},  \nonumber\\ p_l^{\rm cm} &=& \sqrt{(E_l^{\rm cm})^2-m^2_l},\nonumber\\
E_l^{\rm cm} &=& \frac{(s-M^2+m^2_l)}{2 \sqrt{s}}.
\eea
The threshold neutrino energy to create the charged lepton partner is given by
\begin{equation}
E_{\nu_l}^{\rm th} = \frac{(m_l + M_p)^2 - M_n^2}{2 M_n},
\end{equation}
 where $m_l,\; M_p$ and $ M_n$ are the masses of the charged lepton, proton and neutron, respectively.

The differential cross section in the laboratory frame is given by
\begin{equation}
\frac{d\sigma_{tot}(\nu_l)}{dt} = \frac{|\bar{\mathcal{M}}|^2 }{32 \pi E^2_\nu M^2 }f(t),
\end{equation}
where
\begin{equation}
f(t)=\frac{M}{2(E_\nu + M)-\frac{E_{\nu_5}}{p_{\nu_5}^2}(t-m_{\nu_5}^2 + 2E_\nu E_{\nu_5})}.
\end{equation}
We have then verified our analytic calculations  with MadGraph\cite{Alwall:2011uj}, where the nucleon form factor in eq. (\ref{eq:17}) is implemented effectively in the Universal FeynRules Output (UFO) files \cite{Degrande:2011ua}, by fixing  $q^2=M^2_{Z^\prime}$. To measure the goodness of the fit between the BLSM-IS and the measured data, we constructed a 
$\chi^2$ test function as 
\begin{equation}
\chi^2 = \sum_\text{bins}  \frac{(\text{Events}_{\text{th}}-\text{Events}_{\text{ex}})^2}{\delta_{ij}^2},
\end{equation}
with $\delta_{ij}$  the covariance matrix that contains the uncorrelated intrinsic experimental statistic and systematic uncertainties in its diagonal entries.
\begin{figure}
\centering
\includegraphics[scale=0.3]{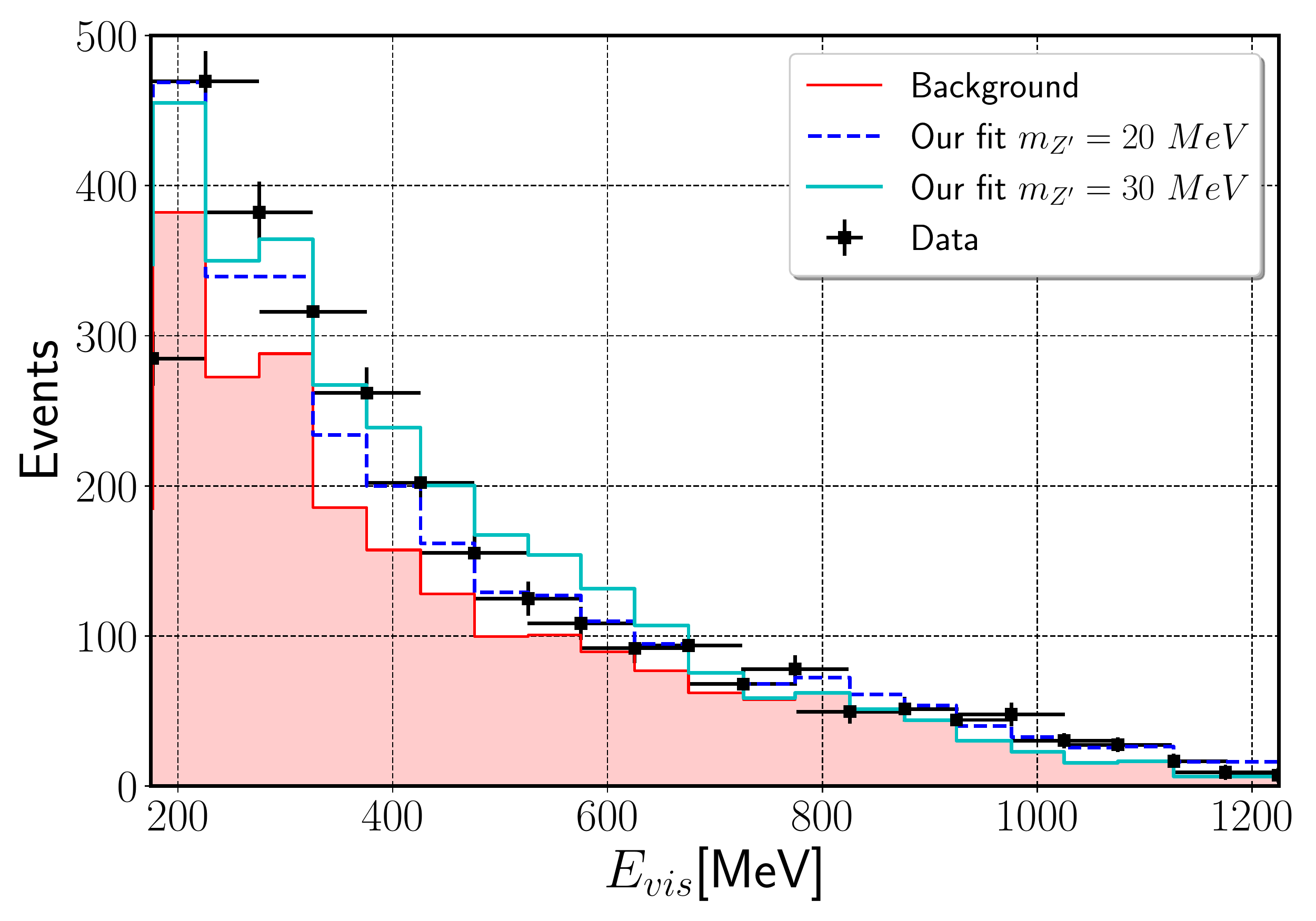}~~
\includegraphics[scale=0.3]{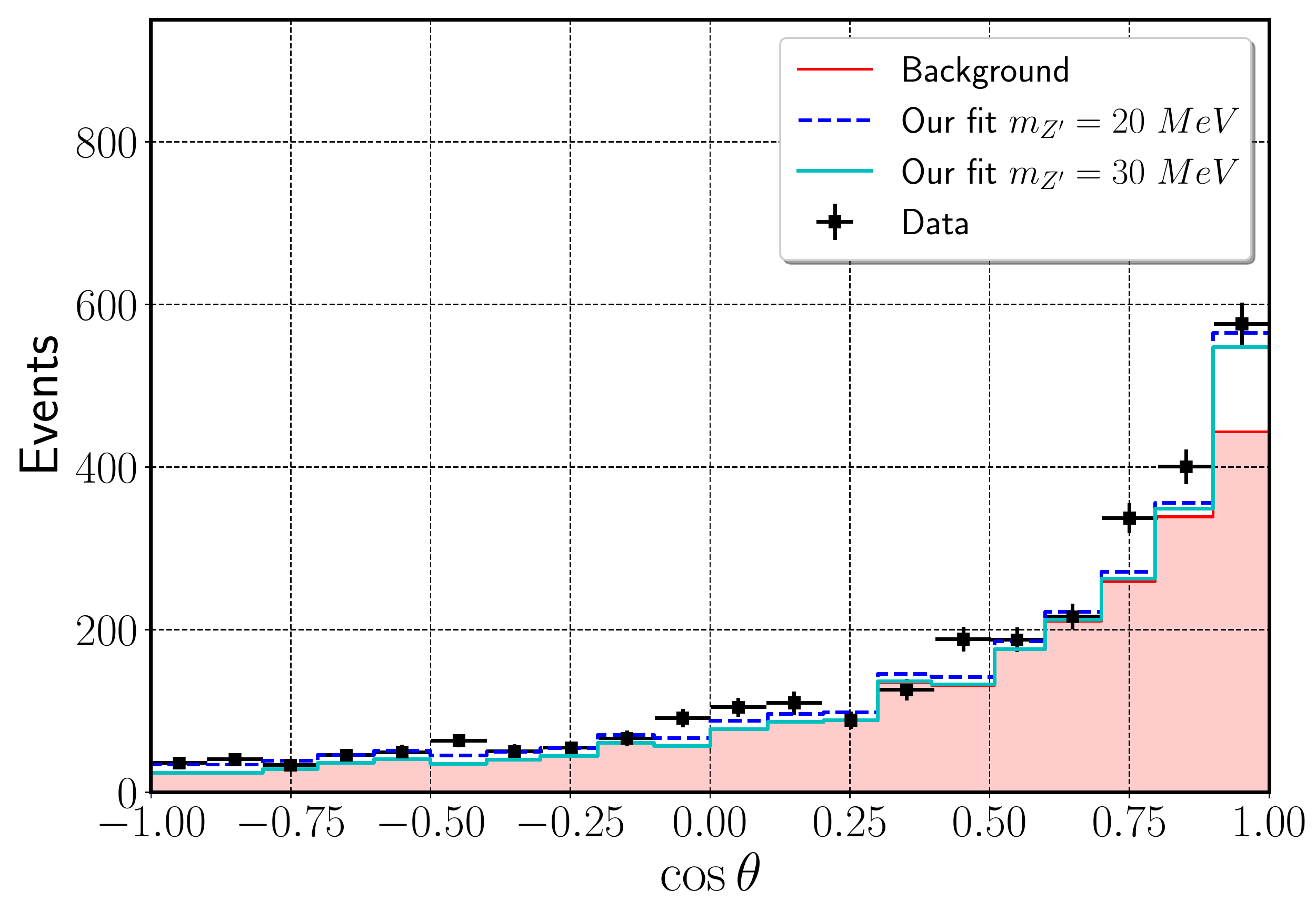}
\caption{Left Panel: The MB electron-like anomalous data and total background events \cite{MiniBooNE:2020pnu} versus the visible energy. Right Panel: Cosin the angle of the final state electron with respect to the beam direction versus the visible energy.  the  Our signal predictions for two benchmark points with $M_{Z^\prime}=20,30$ MeV are presented in dashed blue and solid cyan histograms, respectively.
 }
\label{fig:5}
\end{figure}
Fig. \ref{fig:5}  we show the prediction for two BLSM-IS signals obtained by adopting 
two benchmark points with $M_{Z^\prime}=20,30$ MeV and fixed $g_{(B-L)} =-10^{-4}, \; \tilde{g}=0.2, \; \theta^\prime=3\times 10^{-3}, \; m_{\nu_4} = 60$ MeV and $m_{\nu_5} = 110$ MeV, together with the background and against the data collected by MB which appear anomalous. We find a good agreement between predictions and data up to a $5\sigma$ CL. 
Finally, in fig. \ref{fig:6},  shows the result of the above  fit to the measured MB data extracted from \cite{MiniBooNE:2020pnu} over the $Z^\prime$ mass range  $2-130$ MeV with fixed $m_{\nu_4}, m_{\nu_5}$ and $\theta^\prime$ values while $g_{(B-L)}$ and $\tilde{g}$ have been chosen at their maximal allowed values for the given $M_{Z^\prime}$  (as in fig.\ref{fig:1}). The fit shows that we can reach the $5\sigma$ CL for $Z^\prime$ masses in the range of $15-25$ MeV.

\begin{figure}
\centering
\includegraphics[scale=0.35]{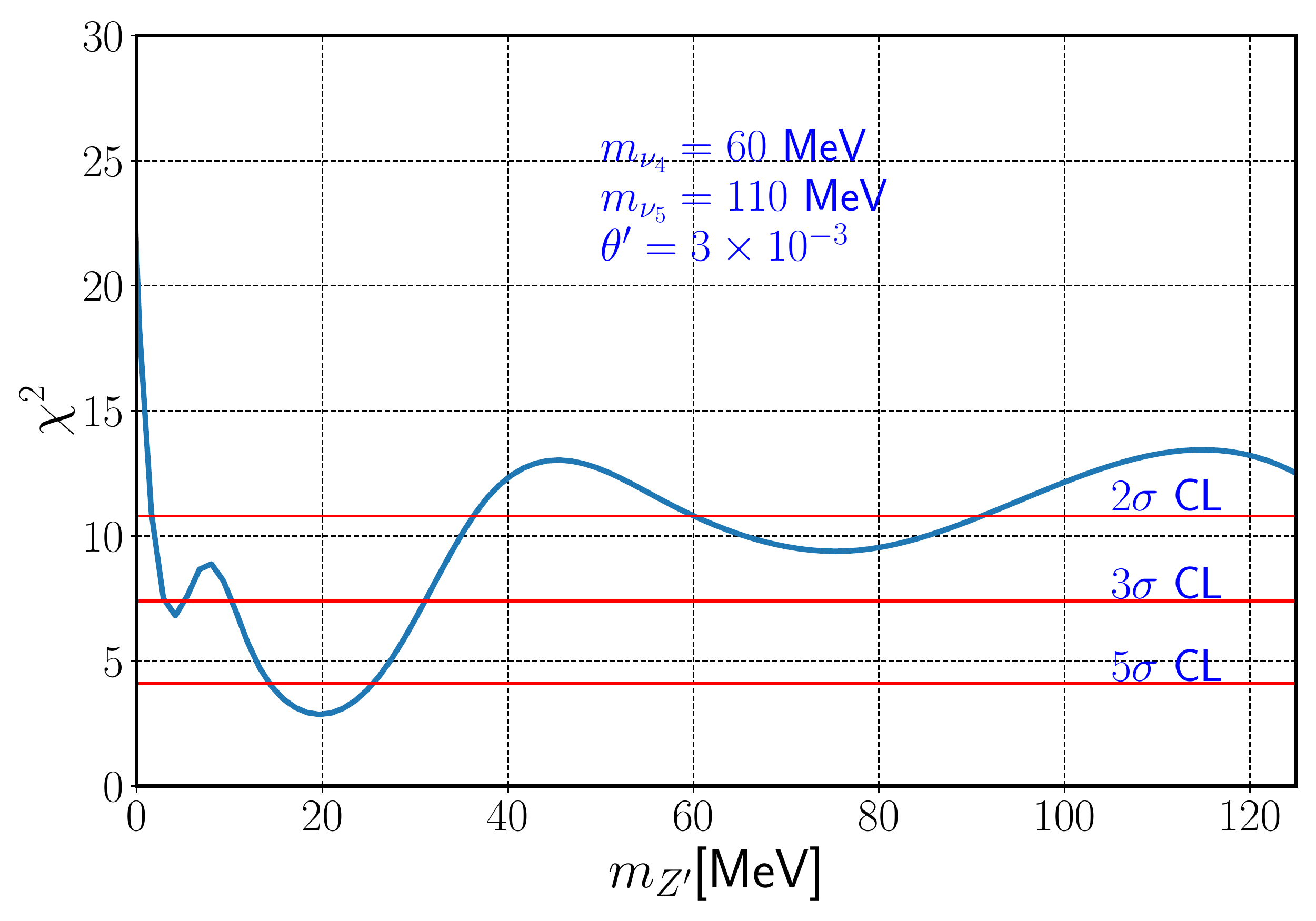}

\caption{ The $\chi^2 $ value of the BLSM-IS  fit  to the MB data versus the $Z^\prime$ mass for $m_{\nu_{5(4)}}=110(60)$ MeV and $\theta'=3\times 10^{-3}$.   Horizontal red lines represents the $2\sigma, 3\sigma$ and $5\sigma$ CL contours. }
\label{fig:6}
\end{figure}

\section{Conclusions}

In summary, in this letter, we have argued that two anomalies presently stemming from non-collider experiments, specifically, in the measurement of
the anomalous magnetic moment of the muon at the  E821 experiment at BNL and the Muon $g-2$ one at FNAL as well as in the study 
 of appearance data in the MB short-baseline neutrino experiment at FNAL, hint at a common explanation relying on some BSM physics that might involve both a light $Z'$ and light neutrinos, all being extremely weakly coupled to the visible sector (so as to being dubbed dark and sterile, respectively).
There is a BSM scenario that can incorporate these new force and matter states in a minimal formulation, thereby being notionally able to explain the aforementioned data sets without invoking an excessing number of new parameters.
This is the so-called BLSM-IS, wherein the SM gauge group is supplemented by an additional, spontaneously broken $U(1)_{B-L}$ invariance, obtained by localising the accidental global $B-L$ conservation of quantum numbers that appears in the SM, in combination with an IS mechanism for neutrino mass generation. The requirement of theoretical self-consistency of this BSM scenario in fact imposes the simultaneous presence of a $Z'$ state following the $B-L$ breaking, which can be made light rather naturally, and of multiple sterile neutrinos, which are {per se} rather light. Herein, we have put the BLSM-IS explanations to the aforementioned data anomalies on firm quantitative grounds. In fact, solutions have been found to both anomalies simultaneously for the following ranges of BLSM-IS parameters: {$M_{Z^\prime}= 15-25$ MeV, $g_{(B-L)} \sim -10^{-4}, \; \tilde{g}\approx 0.2, \; \theta^\prime=3\times 10^{-3}, \; m_{\nu_4} = 60$ MeV and $m_{\nu_5} = 110$ MeV}.

\section*{Acknowledgments}
SM is supported in part through the NExT Institute and  STFC consolidated Grant No.
ST/L000296/1.  AH and AR would like to thank Waleed Abdallah for  fruitful discussions about the MB analysis.  A. Hammad is supported from the Basic Science Research Program through the National Research Foundation of Korea Research Grant No. NRF-2021R1A2C4002551.



\bibliographystyle{h-elsevier}
\bibliography{References}

\end{document}